\documentclass[showpacs,amsfonts,amsmath,amssymb,aps,prl,twocolumn,epsfig.epsf]{revtex4}
\usepackage{bm}
\usepackage{amsfonts}
\usepackage{graphicx}
\begin{document}
\title{Nondispersive two-electron Trojan wave packets}
\author{Matt Kalinski, Loren Hansen, and David Farrelly}
\affiliation{Department of Chemistry and Biochemistry \\
Utah State University, Logan, UT 84322-0300}
\begin{abstract}
We demonstrate the existence of stable non-dispersing two-electron
Trojan-like wave packets in the helium atom in combined magnetic
and circularly polarized microwave fields. These packets follow
circular orbits and we show that they can also exist in quantum
dots. Classically the two electrons follow trajectories which
resemble orbits discovered by Langmuir and which were used in
attempts at a Bohr-like quantization of the helium atom.
Eigenvalues of a generalized Hessian matrix are computed to
investigate the classical stability of these states. Diffusion
Monte Carlo simulations demonstrate the quantum stability of these
two-electron wave packets in the helium atom and quantum dot
helium with an impurity center.
\end{abstract}
\maketitle

The direct manipulation of atoms and ions at the quantum level is currently a
flourishing area of physics \cite{coinsul}.  For example, ion
traps combined with laser cooling techniques have been used to
create new states of matter including ion liquids, Wigner crystals
and Bose-Einstein condensates \cite{corn,wine}. The ability
to manipulate the quantum properties of matter directly, as
exemplified by these advances, is central to the practical development
of nanodevices, e.g., quantum dots and microchip traps
\cite{long}.

Recently it has proved possible to create a trap consisting of a
single atom inside of which the quantum behavior of an electron
can be manipulated \cite{gallag}. In these experiments
\cite{gallag,gallag1} an electron in an excited lithium atom was
localized in a classical orbit almost indefinitely, neither
spreading nor dispersing. This ``classical atom'' was synthesized
by ``tethering'' the electron using a microwave field to which its
motion is phase locked. Potential practical applications of the
technique include Rydberg tagging in molecular spectroscopy and
the preparation of stable antimatter atoms \cite{gallag1}.

Although only recently realized experimentally, the existence of
coherent, non-dispersive one-electron wave packets in Rydberg
atoms was predicted about a decade ago using essentially classical
mechanical arguments \cite{ibb,lee,buchl,lee1}. A peculiar
property of this type of wave packet is that the electron is
localized at an equilibrium which corresponds to an energy {\it
maximum} in the noninertial frame. Such equilibria are similar to
the well known Lagrange equilibrium points $L_4$ and $L_5$ in the
restricted three-body problem of celestial mechanics at which,
e.g., Jupiter's Trojan asteroids are located \cite{murray, lee1}.
Because their classical dynamical origins are so similar, these
wave packets have come to be known as Trojan wave packets
\cite{ibb,lee1}.

Naturally it is interesting to ask if similar wave packets but
involving more than a single electron can be synthesized. In fact,
considerable effort has already been directed towards this goal.
For example, two-electron wave packets in barium atoms \cite{pish}
have been prepared and provide a powerful tool in the study of
electron correlation in atoms \cite{stroud}. However, while these
wave packets are prepared initially in localized radial states
using laser pulses they eventually disperse, e.g., through a
collision near the nucleus. In another approach a particular type
of dynamical stabilization is used to generate nondispersing wave
packets in two-electron atoms. In this case it is
necessary that one of the electrons is well localized relatively
close to the nucleus \cite{stroud}.

Here we report the existence of coherent, two-electron,
non-dispersing wave packets in helium-like atoms (or quantum dot
helium) which are true analogs of the one-electron Trojan wave
packets; both electrons follow classical orbits in direct analogy
with the states prepared experimentally in Refs.
\cite{gallag,gallag1}. Unlike the one-electron Trojan problem,
this system is a genuine quantum three-body problem and thus
represents a more direct analogy with the classical restricted
three-body problem. Stable two-electron equilibria are produced
through the simultaneous application of combined circularly
polarized (C.P.) electromagnetic and magnetic fields to the helium
atom or to quantum dot helium with an impurity center
\cite{ashoori}. The equilibria so produced are stable over broad
ranges of field parameters and the wave packets are, therefore
true non-dispersing coherent-like states.

We demonstrate that, in the two-dimensional limit, e.g., as in
quantum dot helium, it is possible to create a variety of stable
two-electron, non-dispersive, Trojan wave packets. However, there
exists only a single stable three-dimensional configuration in the
helium atom itself. These latter wave packets are actually
two-electron examples of the electronic coherent states sought
after by Schr{\"o}dinger \cite{schr} in the hydrogen atom. In
particular these wave packets follow the ``double-circle" orbits
originally discovered by Langmuir \cite{lang} in the helium atom.
We also present diffusion Monte Carlo simulations which confirm
directly the quantum stability of these states. Further, we show
that it should be possible to produce certain of these states
directly in quantum dots using field strengths which are currently
accessible in the laboratory.

In a coordinate system rotating with the C.P. field and
assuming an infinite nuclear mass the Hamiltonian for
the helium atom interacting with a C.P.
field and a magnetic field perpendicular to the plane
of polarization is, in atomic units,
\begin{eqnarray}
H=H_1+H_2+{1 \over {r_{12}}}
\end{eqnarray}
where,
\begin{eqnarray}
H_{i}&=&{{{\bf p}_i} ^2 \over 2} - {2 \over {r_i}} - (\omega \pm {1 \over 2})(x_i p_{y_i}-y_i p_{x_i}) \nonumber \\
&+& {1 \over 8}(x_i^2+y_i^2) + {\cal E} x_i
\end{eqnarray}
\begin{figure}
\centering
\vspace{10mm}
\hspace{-20mm}
\includegraphics[scale=0.45,angle=0,width=0.45\textwidth]{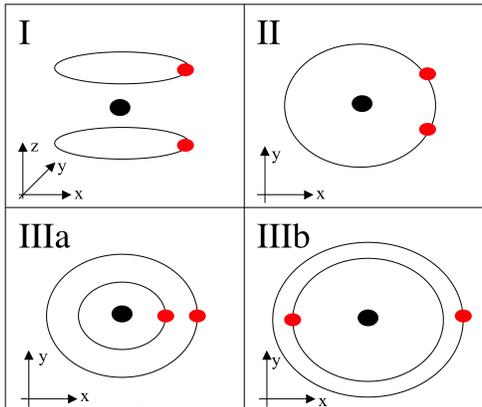}
\vspace{-50mm}
\caption{Possible rotating configurations of two electrons in a helium atom
or quantum-dot helium
in combined magnetic and CP fields. Type I: Langmuir configuration,
type II: transverse configuration, type III: collinear configurations
(a and b).}
\label{fig1}
\end{figure}
\noindent with $i =1,2$; ${r}_i = \sqrt{x_i^2+y_i^2+z_i^2}$ and
${\bf p}_i = (p_{x_i}, p_{y_i}, p_{z_i})$ are the coordinate and
momentum vectors of each electron; $\omega$ and ${\cal E}$ are the
scaled C.P. field frequency and scaled field strength,
respectively \cite{units}; the cyclotron frequency in scaled units
is $\pm {1}$. With a re-interpretation and re-scaling of
parameters this is also the Hamiltonian of quantum dot helium with
an off-axial impurity centre in a magnetic field \cite{lee2}.
Equilibria of the classical Hamiltonian can most easily be found
by constructing a zero-velocity surface (ZVS) \cite{murray}, which
shares some properties with a potential energy surface (PES) but
contains addition terms due to  the centrifugal and Lorentz forces
\cite{murray, lee1}. An extended discussion of the construction of
the ZVS in atomic systems is given in \cite{lee1,contemp} -
briefly the ZVS is obtained by re-writing the Hamiltonian in terms
of velocities rather than momenta. This gives
\begin{eqnarray}
H_{i}&=&\sum_k {1 \over 2} {\dot q_k}^2-{1 \over r_i} + {\cal E} x_i -{\omega (\omega \pm {1 \over 2}) }({x_i}^2+{y_i}^2)
\end{eqnarray}
\noindent where $q_k = x,y,z; k=1,2,3$. Setting the velocities to zero produces the ZVS which, in the
present case, has the form
\begin{eqnarray}
V_{ZVS}= V_{ZVS_1}+V_{ZVS_2} +{1/r_{12}}
\end{eqnarray}
\begin{eqnarray}
V_{{ZVS}_i}=-{1 \over r_i} + {\cal E} x_i -{\omega (\omega \pm {1 \over 2}) }({x_i}^2+{y_i}^2)
\end{eqnarray}
The ZVS reduces to a true PES when $\omega=1/2$ , i.e., when the
paramagnetic term in eq. (2) vanishes. As with a PES equilibria of
the motion are then obtained as extrema of the ZVS (maxima, minima
and saddles) \cite{lee1,contemp}.

We find three possible types of equilibrium point whose stability
depends on the particular values of the parameters. Type I: This
configuration, illustrated in Fig. 1 (I), corresponds to the
unstable ``two-circle" orbits discovered by Langmuir \cite{lang}
in which the (classical) electrons occupy parallel orbits located
above and below the plane of the nucleus with longitudes of
$±30^o$. Stabilization is possible only through the application of
external fields. Type II: Illustrated in Fig. 1 (II), this
configuration has the two electrons orbiting in the same orbit in
a plane containing the nucleus. The angle subtended by the
electrons at the nucleus can be arbitrary. This configuration can
only be stable in two spatial dimensions e.g., a quantum dot. Type
III: This configuration, of which two variants are possible - IIIa
and IIIb in Fig. 1 - is unstable in the
helium atom itself but stable in 2-dimensional quantum dot helium
where it appears as variant IIIa. Both electrons lie in the plane
of polarization and, remarkably, lie on the same side of the
nucleus.

The equilibrium corresponding to the Langmuir configuration has
the geometry of an equilateral triangle; i.e., the three particles form
an equilateral triangle whose sides are of length $a$ (see Fig. 1 a and Fig.
2). The relationship between the field parameters and the size of
this triangle is defined by the following cubic in $a$;
\begin{eqnarray}
{(\omega^2 \pm \omega)} {a^3 \over 2} + {{\sqrt{3}\over 3}}{\cal E} a^2 - 1  = 0\\ \nonumber
\end{eqnarray}
\begin{figure}
\centering
\hspace{-20mm}
\vspace{5mm}
\includegraphics[scale=0.45,angle=0,width=0.45\textwidth]{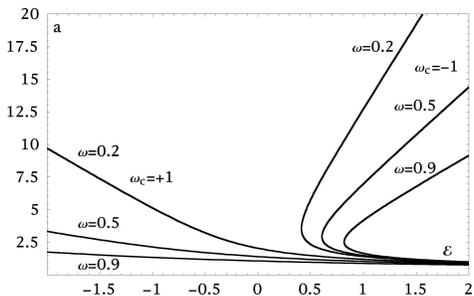}
\vspace{-50mm} \caption{Dependence of the length of the sides of
the equilateral triangle described by the three charges in the
rotating frame on the strength of the CP field for various
rotation frequencies. Note that for $\omega_c=-1$ the system is
bistable and only the larger solution corresponds to the stable
trajectory. When $\omega_c=-1$ there is no critical field and only
a single solution \cite{units}.} \label{fig3}
\end{figure}
Note that the presence of external fields does not change the
geometry of the Langmuir configuration. Figure 2 shows the
dependence of $a$ on the electric field strength for selected
frequencies $\omega$. We perform an extended stability analysis of
the system around these trajectories, i.e. we study  small
oscillations \cite{goldstein}, in the rotating frame. The generalized Hessian matrix (stability or
monodromy matrix) for a system of $n$-particles in the rotating
frame  can be written as 
\begin{eqnarray}
{\bf S}={\bf H}_e {\bf R}
\end{eqnarray}
where
\begin{eqnarray}
{\bf R}_{a_i b_i}=sgn(a_i)\delta_{a_i,b_i}
\end{eqnarray}
where $sgn(a_i)=1, a_i=p_i$,
$sgn(a_i)=-1, a_i=x_i$
and $H_e$ is the phase space defined Hessian matrix
\begin{eqnarray}
({{\bf H}_e})_{a_i b_i}={{{\partial^2 H}_{osc} \over {\partial a_i \partial b_i}}}
\end{eqnarray}
\begin{figure}
\centering
\hspace{-20mm} \includegraphics[scale=0.45,angle=0,width=0.45\textwidth]{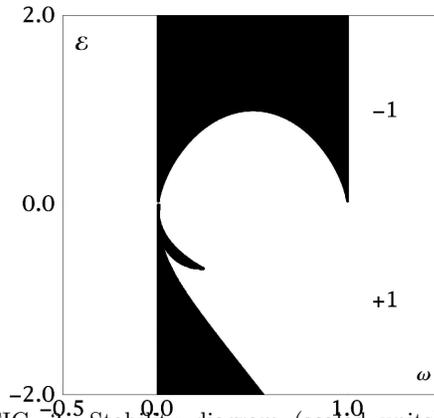}
\vspace{-30mm}
\caption{Stability diagram (scaled units) of the disturbed Langmuir trajectory
for anti-centrifugal Lorentz force (upper half, $\omega_c=-1$) and
co-centrifugal (lower half, $\omega_c=+1$) as a function of scaled rotation
frequency and the electric field. The black regions corresponds to stable trajectories. Note that no magnetic 
field  $\omega_c<\omega$ can lead to stability in  case $-1$ and magnetic field weaker than rotation $\omega_c<\omega$ is 
permitted to obtain stability in case $+1$. This  suggests a maximum of the ZVS and Paul trap-like (Trojan) stabilization.}
\label{fig2}
\end{figure}
$a_i=q_i, p_i$,
$b_i=q_i, p_i$
and $H_{osc}$ is the harmonic Hamiltonian around the orbit \cite{ibb,hell}.

The generalized Hessian matrix can be written explicitly as
\begin{center}
\hspace{17mm}
${\bf S}_{ij}=$\( \left [ \begin{array}{cccc}
{\bf A} & {\bf 0} &   {\bf C}_{11} & {\bf C}_{12} \\
{\bf 0} &  {\bf A} &   {\bf C}_{21} & {\bf C}_{22}     \\ 
{\bf B} & {\bf 0}  &   {\bf A}  &  {\bf 0}   \\
{\bf 0} & {\bf B}  &   {\bf 0}  & {\bf A}
\end{array}  \right ] \)
$\:\:\:\:\:\:\:\:\:\:\:\:\:\:\:\:\:\:\:\:\:\:\:\:$(10)
\end{center}
Note that the right upper block is the normal Hessian matrix of the
stationary mechanical system $C_{ij}$ \cite{goldstein}
and
\begin{center}
\hspace{10mm}
${\bf A}=$\( \left [ \begin{array}{ccc}
0 & {\omega \pm {1 \over 2}} & 0 \\
-\omega \mp {1 \over 2} & 0 & 0 \\
0 & 0 & 0 \\
\end{array} \right ] \),
${\bf B}=$\( \left [  \begin{array}{ccc}
1 & 0 & 0  \\
0 & 1 & 0  \\
0 & 0 & 1
\end{array}  \right ] \)
$\:\:\:\:\:$(11)
\end{center}
Figure 3 shows the stability island as a function of
scaled coordinates \cite{lee} for the cases $\omega_c = 1$
(lower half)
and $\omega_c = -1$.
Remarkably, unlike for Trojan wavepackets both the circularly polarized
field and the magnetic field are necessary to stabilize the Langmuir
trajectories. 
One may notice an interesting v-shaped valley which is similar to  the ``negative mass" case of Ref. \cite{mejoe}.

The quantum wave function for the monodromy matrix for the stable trajectory \cite{hell} may be written \cite{ibb}
\begin{eqnarray}
\:\:\:\:\:\:\:\:\:\:\:\:\:\:\:\:\:\:\:\:\:\:\:\:\:\:\:\:\:\:\:\:\psi=e^{-\sum_{ij} A_{ij} x_i x_j} \nonumber \:\:\:\:\:\:\:\:\:\:\:\:\:\:\:\:\:\:\:\:\:\:\:\:\:\:\:\:\:\:\:\ (12)
\end{eqnarray}
where $A_{ii}$ are always real and $A_{ij}$, $i \ne j$  imaginary to guarantee normalizability of the quantum wavefunction in the rotating frame.
Because  the wavefunction is localized around the stable point in the
configuration plane the  contraction (integration)  of diagonal of the density matrix $\psi^{*}(x) \psi(x)$ over a single 
electron variable leads to the single electron density localized around
electron equilibria and corresponds to a nondispersing two-electron wavepacket moving in the laboratory frame.
The detailed calculations are quite cumbersome and will be given elsewhere.
As one can see from Fig. 3 the system never stabilizes for magnetic fields weaker than the electric field frequency  for the case of anti-centrifugal
Lorentz force and may stabilize for such fields for faster rotations
for the  co-centrifugal Lorentz force.

In order to check for the existence of the  quantum states themselves we solved the time dependent 
Schr{\"o}dinger equation using the Diffusion Monte Carlo method \cite{diff,diff1}.
Fig. 4 shows density plots of the wave function of each electron.
The probability density is clearly localized around the
equilibrium points. The wavepackets are actually approximate eigenstates of the Hamiltonian
in the rotating frame and, therefore, do not disperse in the inertial frame.
These eigenstates are localized around a classical equilibrium point whose
local spectrum is an almost-harmonic ladder of coherent states. That is, the
eigenstate of the atom in the fields itself behaves as a wavepacket evolving
according to the classical equations of motion. For the large field values
employed here the Langmuir wavepackets actually correspond to the ground state
of eq. (2). This is similar to the case of the original Trojan wavepackets \cite{ibb}
which are the ground states of the locally harmonic Hamiltonian obtained by expansion around
the Trojan equilibria which are, in that case, energy maxima \cite{lee}. We neglect finite nucleus mass effects here but on the timescales
of interest, as in the case of one-electron Trojan wavepackets,
these effects will be negligible \cite{schmeltz}. 


\begin{figure}
\centering
\hspace{-20mm}
\includegraphics[scale=0.33,angle=0,width=0.33\textwidth]{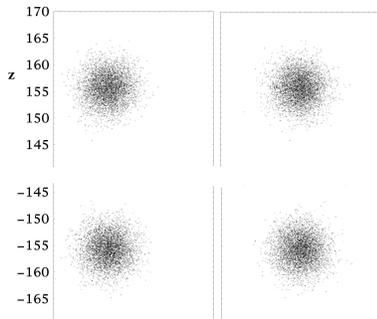}
\vspace{-30mm}
\caption{Diffusion Monte Carlo simulations of Langmuir wavepackets in the rotating frame for
$\Omega_c=0.0370$ a.u., ${\cal E} = 0.1235$a.u., and
$\Omega=\Omega_c/2$. The density of points reflects the wavefunction itself not probability density \cite{lee2,diff}.
The wavepackets are approximate eigenstates of the Hamiltonian
in the rotating frame and follow a circular orbit in the inertial frame.}
\label{fig4}
\end{figure}

In conclusion we have demonstrated that Langmuir trajectories in
magnetic and circularly polarized fields are stabilizing for
certain  parameter regions when the magnetic field is parallel to
the rotation axis and the C.P. field is perpendicular.
Corresponding quantum states exist. This  is also true for
two-dimensional quantum dot helium. These regions are not possible
to predict from purely analytical considerations and require
extensive numerical searches in parameter space. For example, for
a quantum dot of radius  100 nm (Type IIIa configuration) the
equilibria are at 62.72 nm and 98.00 nm from the nucleus. To
achieve this configuration a magnetic field of 5 T is  applied and 
the impurity with the effective charge $Z_{eff}$=0.008 {\it e}  displaced from the center of the parabolic dot by 98 nm generates effective 
``microwave" field of frequency 548 GHz and strength  4.593 kV/m.

 \acknowledgements
Financial Support by the (US) National Science foundation and the
Petroleum Research Fund, administered by the American Chemical
Society is gratefully acknowledged.



\begin{references}
\bibitem{units} We scale the rotation frequency $\Omega$ to the cyclotronic frequency $\Omega_c$
namely $\omega=\Omega/\Omega_c$ and the cyclotronic frequency to itself,
$\omega_c=\pm 1= |\Omega_c|/\Omega_c$. Cartesian coordinates and electric field are scaled accordingly.
\bibitem{diff1} Within the Diffusion Monte Carlo method one solves the
time dependent imaginary time Schr{\"o}dinger equation using an
ensemble of particles whose  density is the quantum wavefunction
itself. The solution converges to the ground state (or excited
states if appropriate nodal boundary conditions are imposed).
\end{references}
\end{document}